\documentclass[conference]{IEEEtran}
\usepackage{multicol}
\usepackage{lipsum,graphicx}
\usepackage{float}
\usepackage{epstopdf}
\usepackage{ifthen}
\usepackage[latin1]{inputenc}

\usepackage{amssymb}
\usepackage[cmex10]{amsmath}
\usepackage{dcolumn}
\usepackage[nolist]{acronym}

\DeclareGraphicsExtensions{.pdf,.jpeg,.png,.eps}
\usepackage{booktabs}

\usepackage[caption=false,font=footnotesize]{subfig}
\usepackage{comment}
\usepackage{color}
\usepackage[table]{xcolor}
\usepackage{cite}

\usepackage{setspace}
\usepackage{mathrsfs}          
\usepackage{amsmath, amssymb, amsthm}
\usepackage{amsfonts}
\usepackage{mathrsfs}
\usepackage{multirow}        
\usepackage{framed}
\usepackage{psfrag}
\usepackage{graphicx}
\usepackage{textcomp, gensymb} 
\usepackage{algpseudocode} 

\newcolumntype{d}[1]{D{.}{\cdot}{#1} }

\usepackage{bm}
\usepackage{mathtools}
\usepackage{siunitx}

\newcommand{\PreserveBackslash}[1]{\let\temp=\\#1\let\\=\temp}
\newcolumntype{C}[1]{>{\PreserveBackslash\centering}p{#1}}
\newcolumntype{R}[1]{>{\PreserveBackslash\raggedleft}p{#1}}
\newcolumntype{L}[1]{>{\PreserveBackslash\raggedright}p{#1}}

\newcommand{\figref}[1]{Fig.~\ref{#1}}

\usepackage{lipsum}
\usepackage{mathtools}
\usepackage{cuted}
\usepackage{stmaryrd}

\usepackage{tabularx}
\usepackage{tabulary}
\usepackage{tikz}
\usepackage{pgfplots}
\usepgfplotslibrary{units}
\pgfplotsset{compat=1.18}

\usepackage{algpseudocode}
\usepackage{algorithm}

\begin{document}
	\begin{acronym}
\acro{1G}{first generation}
\acro{2G}{second generation}
\acro{3G}{third generation}
\acro{3GPP}{Third Generation Partnership Project}
\acro{4G}{fourth generation}
\acro{5G}{fifth generation}
\acro{6G}{sixth-generation}
\acro{802.11}{IEEE 802.11 specifications}
\acro{A/D}{analog-to-digital}
\acro{ADC}{analog-to-digital}
\acro{AGC}{automatic gain control}
\acro{AM}{amplitude modulation}
\acro{AoA}{angle of arrival}
\acro{AP}{access point}
\acro{AI}{artificial intelligence}
\acro{AR}{augmented reality}
\acro{ASIC}{application-specific integrated circuit}
\acro{ASIP}{Application Specific Integrated Processors}
\acro{AWGN}{additive white Gaussian noise}
\acro{BB}{base-band}
\acro{BCJR}{Bahl, Cocke, Jelinek and Raviv}
\acro{BER}{bit error rate}
\acro{BFDM}{bi-orthogonal frequency division multiplexing}
\acro{BPSK}{binary phase shift keying}
\acro{BS}{base stations}
\acro{CA}{carrier aggregation}
\acro{CAF}{cyclic autocorrelation function}
\acro{Car-2-x}{car-to-car and car-to-infrastructure communication}
\acro{CAZAC}{constant amplitude zero autocorrelation waveform}
\acro{CB-FMT}{cyclic block filtered multitone}
\acro{CCDF}{complementary cumulative density function}
\acro{CDF}{cumulative density function}
\acro{CDMA}{code-division multiple access}
\acro{CFO}{carrier frequency offset}
\acro{CIR}{channel impulse response}
\acro{CM}{complex multiplication}
\acro{COFDM}{coded-\acs{OFDM}}
\acro{CoMP}{coordinated multi point}
\acro{COQAM}{cyclic OQAM}
\acro{CP}{cyclic prefix}
\acro{CRB}{Cramer-Rao bound}
\acro{CPE}{constant phase error}
\acro{CR}{cognitive radio}
\acro{CRC}{cyclic redundancy check}
\acro{CRLB}{Cram\'{e}r-Rao lower bound}
\acro{CS}{cyclic suffix}
\acro{CSI}{channel state information}
\acro{CSMA}{carrier-sense multiple access}
\acro{CWCU}{component-wise conditionally unbiased}
\acro{D/A}{digital-to-analog}
\acro{D2D}{device-to-device}
\acro{DAC}{digital-to-analog converter}
\acro{DBF}{digital beamforming}
\acro{DC}{direct current}
\acro{DFE}{decision feedback equalizer}
\acro{DFT}{discrete Fourier transform}
\acro{DL}{downlink}
\acro{DMT}{discrete multitone}
\acro{DNN}{deep neural network}
\acro{DSA}{dynamic spectrum access}
\acro{DSL}{digital subscriber line}
\acro{DSP}{digital signal processor}
\acro{DTFT}{discrete-time Fourier transform}
\acro{DVB}{digital video broadcasting}
\acro{DVB-T}{terrestrial digital video broadcasting}
\acro{DWMT}{discrete wavelet multi tone}
\acro{DZT}{discrete Zak transform}
\acro{E2E}{end-to-end}
\acro{eNodeB}{evolved node b base station}
\acro{E-SNR}{effective signal-to-noise ratio}
\acro{EVD}{eigenvalue decomposition}
\acro{EVM}{error vector magnitude}
\acro{FBMC}{filter bank multicarrier}
\acro{FD}{frequency-domain}
\acro{FDA}{frequency diverse array}
\acro{FDD}{frequency-division duplexing}
\acro{FDE}{frequency domain equalization}
\acro{FDM}{frequency division multiplex}
\acro{FDMA}{frequency-division multiple access}
\acro{FEC}{forward error correction}
\acro{FER}{frame error rate}
\acro{FFT}{fast Fourier transform}
\acro{FI}{Fisher information}
\acro{FIR}{finite impulse response}
\acro{FM}		{frequency modulation}
\acro{FMT}{filtered multi tone}
\acro{FO}{frequency offset}
\acro{F-OFDM}{filtered-\acs{OFDM}}
\acro{FPGA}{field programmable gate array}
\acro{FSC}{frequency selective channel}
\acro{FS-OQAM-GFDM}{frequency-shift OQAM-GFDM}
\acro{FT}{Fourier transform}
\acro{FTD}{fractional time delay}
\acro{FTN}{faster-than-Nyquist signaling}
\acro{GFDM}{generalized frequency division multiplexing}
\acro{GFDMA}{generalized frequency division multiple access}
\acro{GMC-CDM}{generalized	multicarrier code-division multiplexing}
\acro{GNSS}{global navigation satellite system}
\acro{GS}{guard symbols}
\acro{GSM}{Groupe Sp\'{e}cial Mobile}
\acro{GUI}{graphical user interface}
\acro{H2H}{human-to-human}
\acro{H2M}{human-to-machine}
\acro{HF}{high frequency}
\acro{HPBW}{half-power beam-width}
\acro{HTC}{human type communication}
\acro{I}{in-phase}
\acro{i.i.d.}{independent and identically distributed}
\acro{IB}{in-band}
\acro{IBI}{inter-block interference}
\acro{IC}{interference cancellation}
\acro{ICI}{inter-carrier interference}
\acro{ICT}{information and communication technologies}
\acro{ICV}{information coefficient vector}
\acro{IDFT}{inverse discrete Fourier transform}
\acro{IDMA}{interleave division multiple access}
\acro{IEEE}{institute of electrical and electronics engineers}
\acro{IF}{intermediate frequency}
\acro{IFFT}{inverse fast Fourier transform}
\acro{IoT}{Internet of Things}
\acro{IOTA}{isotropic orthogonal transform algorithm}
\acro{IP}{internet protocole}
\acro{IP-core}{intellectual property core}
\acro{ISDB-T}{terrestrial integrated services digital broadcasting}
\acro{ISDN}{integrated services digital network}
\acro{ISI}{inter-symbol interference}
\acro{ITU}{International Telecommunication Union}
\acro{IUI}{inter-user interference}
\acro{JCnS}{joint communication and sensing}
\acro{KDE}{kernel density estimate}
\acro{LAN}{local area netwrok}
\acro{LLR}{log-likelihood ratio}
\acro{LMMSE}{linear minimum mean square error}
\acro{LNA}{low noise amplifier}
\acro{LO}{local oscillator}
\acro{LOS}{line-of-sight}
\acro{LoS}{line of sight}
\acro{LP}{low-pass}
\acro{LPF}{low-pass filter}
\acro{LS}{least squares}
\acro{LTE}{long term evolution}
\acro{LTE-A}{LTE-Advanced}
\acro{LTIV}{linear time invariant}
\acro{LTV}{linear time variant}
\acro{LUT}{lookup table}
\acro{M2M}{machine-to-machine}
\acro{MA}{multiple access}
\acro{MAC}{multiple access control}
\acro{MAP}{maximum a posteriori}
\acro{MAE}{mean absolute error}
\acro{MC}{multicarrier}
\acro{MCA}{multicarrier access}
\acro{MCM}{multicarrier modulation}
\acro{MCS}{modulation coding scheme}
\acro{MF}{matched filter}
\acro{MF-SIC}{matched filter with successive interference cancellation}
\acro{MIMO}{multiple-input multiple-output}
\acro{MISO}{multiple-input single-output}
\acro{ML}{maximum likelihood}
\acro{MLD}{maximum likelihood detection}
\acro{MLE}{maximum likelihood estimator}
\acro{MMSE}{minimum mean squared error}
\acro{MRC}{maximum ratio combining}
\acro{MS}{mobile stations}
\acro{MSE}{mean squared error}
\acro{MSK}{Minimum-shift keying}
\acro{MSSS}[MSSS]	{mean-square signal separation}
\acro{MTC}{machine type communication}
\acro{MU}{multi user}
\acro{MVDR}{minimum variance distortionless response}
\acro{MVUE}{minimum variance unbiased estimator}
\acro{NEF}{noise enhancement factor}
\acro{NLOS}{non-line-of-sight}
\acro{NMSE}{normalized mean-squared error}
\acro{NOMA}{non-orthogonal multiple access}
\acro{NPR}{near-perfect reconstruction}
\acro{NRZ}{non-return-to-zero}
\acro{OU}{Ornstein-Uhlenbeck}
\acro{OFDM}{orthogonal frequency division multiplexing}
\acro{OFDMA}{orthogonal frequency division multiple access}
\acro{OMP}{orthogonal matching pursuit}
\acro{OOB}{out-of-band}
\acro{OQAM}{offset quadrature amplitude modulation}
\acro{OQPSK}{offset quadrature phase shift keying}
\acro{OTA}{Over-the-Air}
\acro{OTFS}{orthogonal time frequency space}
\acro{PA}{power amplifier}
\acro{PAM}{pulse amplitude modulation}
\acro{PAPR}{peak-to-average power ratio}
\acro{PB}{pass-band}
\acro{PC-CC}{parallel concatenated convolutional code}
\acro{PCP}{pseudo-circular pre/post-amble}
\acro{PD}{probability of detection}
\acro{pdf}{probability density function}
\acro{PDF}{probability distribution function}
\acro{PDP}{power delay profile}
\acro{PFA}{probability of false alarm}
\acro{PFD}{phase-frequency detector}
\acro{PHY}{physical layer}
\acro{PIC}{parallel interference cancellation}
\acro{PLC}{power line communication}
\acro{PLL}{phase-locked loop}
\acro{PMF}{probability mass function}
\acro{PN}{pseudo noise}
\acro{ppm}{parts per million}
\acro{PRB}{physical resource block}
\acro{PRB}{physical resource block}
\acro{PSD}{power spectral density}
\acro{Q}{quadrature-phase}
\acro{QAM}{quadrature amplitude modulation}
\acro{QoS}{quality of service}
\acro{QPSK}{quadrature phase shift keying}
\acro{R/W}{read-or-write}
\acro{RAM}{random-access memmory}
\acro{RAN}{radio access network}
\acro{RAT}{radio access technologies}
\acro{RC}{raised cosine}
\acro{RF}{radio frequency}
\acro{rms}{root mean square}
\acro{RMS}{root mean square}
\acro{RMSE}{root mean square error}
\acro{RSSI}{received signal strength indicator}
\acro{RRC}{root raised cosine}
\acro{REF}{reference}
\acro{RW}{read-and-write}
\acro{RX}{receiver}
\acro{SC}{single-carrier}
\acro{SCA}{single-carrier access}
\acro{SC-FDE}{single-carrier with frequency domain equalization}
\acro{SC-FDM}{single-carrier frequency division multiplexing}
\acro{SC-FDMA}{single-carrier frequency division multiple access}
\acro{SD}{sphere decoding}
\acro{SDD}{space-division duplexing}
\acro{SDMA}{space division multiple access}
\acro{SDR}{software-defined radio}
\acro{SDW}{software-defined waveform}
\acro{SEFDM}{spectrally efficient frequency division multiplexing}
\acro{SE-FDM}{spectrally efficient frequency division multiplexing}
\acro{SER}{symbol error rate}
\acro{SIC}{successive interference cancellation}
\acro{SINR}{signal-to-interference-plus-noise ratio}
\acro{SIR}{signal-to-interference ratio}
\acro{SISO}{single-input, single-output}
\acro{SMS}{Short Message Service}
\acro{SNR}{signal-to-noise ratio}
\acro{STC}{space-time coding}
\acro{STFT}{short-time Fourier transform}
\acro{STO}{sample-time-offset}
\acro{SU}{single user}
\acro{SVD}{singular value decomposition}
\acro{TX}{transmitter}
\acro{TD}{time-domain}	
\acro{TDD}{time-division duplexing}
\acro{TTD}{true-time delay}
\acro{TDMA}{time-division multiple access}
\acro{TFL}{time-frequency localization}
\acro{TO}{time offset}
\acro{ToA}{time of arrival}
\acro{TDoA}{time difference of arrival}
\acro{TS-OQAM-GFDM}{time-shifted OQAM-GFDM}
\acro{UE}{user equipment}
\acro{UFMC}{universally filtered multicarrier}
\acro{UL}{uplink}
\acro{ULA}{uniform linear array}
\acro{US}{uncorrelated scattering}
\acro{USB}{universal serial bus}
\acro{USRP}{universal software radio peripheral}
\acro{UW}{unique word}
\acro{VLC}{visible light communications}
\acro{VR}{virtual reality}
\acro{VCO}{voltage-controlled oscillator}
\acro{WCP}{windowing and \acs{CP}}	
\acro{WHT}{Walsh-Hadamard transform}
\acro{WiMAX}{worldwide interoperability for microwave access}
\acro{WLAN}{wireless local area network}
\acro{W-OFDM}{windowed-\acs{OFDM}}	
\acro{WOLA}{windowing and overlapping}	
\acro{WSS}{wide-sense stationary}
\acro{ZCT}{Zadoff-Chu transform}
\acro{ZF}{zero-forcing}
\acro{ZMCSCG}{zero-mean circularly-symmetric complex Gaussian}
\acro{ZP}{zero-padding}
\acro{ZT}{zero-tail}
\acro{URLLC}{ultra-reliable low-latency communications}

\acro{HSI}{human system interface}
\acro{HMI}{human machine interface}
\acro{VR} {visual reality} 
\acro{AGV}{automated guided vehicles}
\acro{MEC}{multiaccess edge cloud}
\acro{TI} {tactile Internet}
\acro{IMT}{ international mobile telecommunications}
\acro{GN}{gateway node}
\acro{CN}{control node}
\acro{NC}{network controller}
\acro{SN}{sensor node}
\acro{AN}{actuator node}
\acro{HN}{haptic node}
\acro{TD}{tactile devices}
\acro{SE}{supporting engine}
\acro{AI}{artificial intelligence}
\acro{TSM}{tactile service manager}
\acro{TTI}{transmission time interval}
\acro{NR}{new radio}
\acro{SDN}{software defined networking}
\acro{NFV}{ network function virtualization}
\acro{CPS}{cyber-physical system}
\acro{TSN}{Time-Sensitive Networking}
\acro{FEC}{forward error correction}
\acro{STC}{space-time  coding}
\acro{HARQ}{hybrid automatic repeat request}
\acro{CoMP} {Coordinated multipoint}
\acro{HIS}{human system interface }
\acro{RU}{radio unit}
\acro{CU}{central unit}
\acro{AoD} {angle of departure}
\end{acronym}
    \title{Proof of Concept: Local TX Real-Time Phase Calibration in MIMO Systems}
	
	\author{
		\IEEEauthorblockN{
			Carl Collmann, Ahmad Nimr, Gerhard Fettweis 
			}
			
		\IEEEauthorblockA{
		Vodafone Chair Mobile Communications Systems, Technische Universit\"{a}t Dresden, Germany\\ \small\texttt{\{carl.collmann, ahmad.nimr,  gerhard.fettweis\}@tu-dresden.de}\\
		}
		}
	\maketitle
	\IEEEpeerreviewmaketitle
	
\begin{abstract}
Channel measurements in \ac{MIMO} systems hinge on precise synchronization.
While methods for time and frequency synchronization are well established, maintaining real-time phase coherence remains an open requirement for many \ac{MIMO} systems.
Phase coherence in \ac{MIMO} systems is crucial for beamforming in digital arrays and enables precise parameter estimates such as Angle-of-Arrival/Departure.
This work presents and validates a simple local real-time phase calibration method for a digital array.
We compare two different approaches, instantaneous and smoothed calibration, to determine the optimal interval between synchronization procedures.
To quantitatively assess calibration performance, we use two metrics: the average beamforming power loss and the \ac{RMS} cycle-to-cycle jitter.
Our results indicate that both approaches for phase calibration are effective and yield \ac{RMS} of jitter in the $\SI{2.1}{ps}$ to $\SI{124}{fs}$ range for different \ac{SDR} models.
This level of precision enables coherent transmission on commonly available \ac{SDR} platforms, allowing investigation on advanced \ac{MIMO} techniques and transmit beamforming in practical testbeds.
\end{abstract}

\begin{IEEEkeywords}
software-defined radio, multiple-input multiple-output, radio frequency transceiver
\end{IEEEkeywords}
 
	\acresetall
\section{Introduction}\label{sec:introduction}

Phase noise in \ac{MIMO} communication systems significantly degrades data throughput \cite{6902790_larsson} and in the context of \ac{JCnS} limits the accuracy of parameter estimates \cite{fanliu_9737357}.
The compensation of phase noise at the receiver side is well established.
For example the 3GPP standard \cite{3gpp_nr_ptrs_chapter_7_4} offers the phase tracking reference signal (PT-RS) to compensate for the \ac{CPE} induced by phase noise.
However to enable transmit beamforming in a digital array, coherence between \ac{RF} chains feeding the transmitting antenna array has to be assured.

There are several known calibration approaches in the literature, each with limitations.
Reciprocity calibration \ac{OTA} with distributed sensor nodes such as \cite{9173801_prager} and \cite{9554042_cubesat} lacks a shared phase reference, making it unsuitable for coherent combination of the signals from multiple transmitters.
In \cite{10438456_leinonen} a reciprocity calibration \ac{OTA} scheme is employed where both transmit and receive arrays are connected to a common measurement device.
This wired connection requires the transmitter and receiver to be co-located, making the approach unsuitable for mobile communication scenarios.
The authors in \cite{10130056_cao} present \ac{OTA} reciprocity calibration in \ac{MIMO} systems, where all distributed radio units are connected to a central controller that performs synchronization by pre-coding with phase estimates. 
A limitation in this work is that the radio units have no method of compensating the \ac{LO} phase drifts internally, which imposes a significant overhead on the \ac{OTA} calibration.
Compounding these challenges, the often neglected residual transmit side calibration errors, can substantially degrade achievable MIMO throughput \cite{5456453_studer}.

In this work, we address these limitations by proposing and validating a simple, local method for real-time phase calibration of transmit RF chains.
Our approach uses a dedicated reference RF chain at the transmitter to receive calibration signals (PT-RS) from each transmit chain in a \ac{TDMA} scheme.
The transmit-controller estimates the phase of each chain relative to this reference and applies precoding to achieve coherent transmission in passband.
The calibration procedure is performed in periodic intervals to continuously track \ac{LO} drift, to directly address the transmit side phase impairments as highlighted in \cite{5456453_studer}.


The paper is organized as follows.
Section \ref{sec:system_model} introduces the system model for local transmitter calibration and modeling of phase noise processes for \ac{VCO} and \ac{PLL}.
Section \ref{sec:impairment_model} presents measurement results and validates the calibration approaches.
Section \ref{sec:evaluation} explores the effect of different calibration intervals and their impact on the achievable beamforming gain.
Finally, the paper is concluded in section \ref{sec:conclusions} with key results.





\section{System Model}\label{sec:system_model}


\subsection{Local Calibration of TX Array}

\begin{figure}[tb]
    	\centering
    	\includegraphics[width=\linewidth]{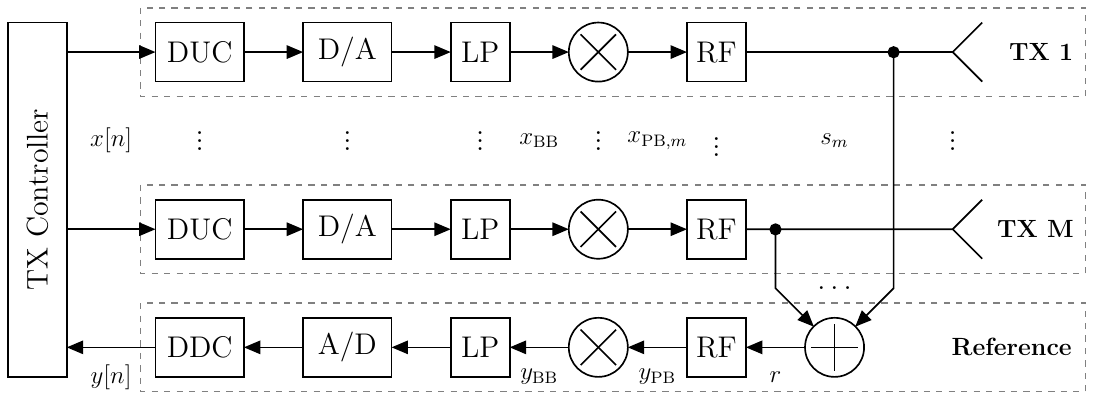}
        \caption{Setup for phase calibration of TX array, comparison to reference.}
    	\label{fig:TX_sync_setup}
\end{figure}

\begin{figure}[tb]
    	\centering
    	\includegraphics[width=0.9\linewidth]{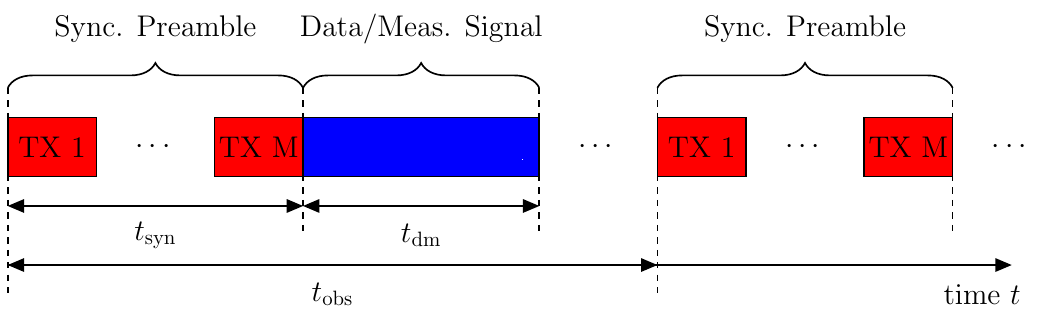}
        \caption{Phase calibration scheme, periodic transmission of synchronization preamble, observed signal at transmitter reference.}
    	\label{fig:TX_sync_scheme}
\end{figure}


Our objective is to calibrate the phase of each element in a transmitting digital antenna array with $M$ elements, as illustrated in \figref{fig:TX_sync_setup}.
To achieve this objective, the phases of the transmitting chains must be compared to a common reference.
Because the transmitting chains and reference RF chain are co-located, we assume a common trigger signal (e.g., 1-pulse-per-second) to align the sampling clocks and a common $\SI{10}{MHz}$ reference are available.
While the sampling clocks are considered as ideal, the synthesized carrier signals exhibit phase noise and corresponding drift.
Further impairments considered are the phase responses of \ac{RF} chain components and their respective \ac{RF} front ends.

As shown in \figref{fig:TX_sync_scheme}, each TX chain transmits a synchronization signal $x[n]$ of length $N$ in \ac{TDMA} before the data or measurement signal.
The band-limited baseband signal after D/A conversion and low-pass filtering is denoted $x_\text{BB}(t)$, defined over $t \in [0, NT_\text{s}]$.
After up-conversion, the bandpass signal for chain $m$ becomes
\begin{align}
    x_{\text{BP},m}(t) = x_\text{BB}(t) e^{j(2\pi f_{\text{c},m}t + \theta_{\text{OS},m}(t))}.
\end{align}
The synthesized carrier frequency $f_{\text{c},m}$ is variable per chain to model a residual frequency offset.
This small residual \ac{CFO} models a linear phase shift over time, distinct from oscillator phase noise.
The term $\theta_{\text{OS},m}(t)$ refers to the time-dependent phase of the oscillator signal for \ac{RF} chain $m$.
Then the transmitted signal at \ac{RF} chain $m$ in \ac{TDMA} is
\begin{align}
    s_{\text{TX},m}(t) = x_{\text{BP},m}(t - mNT_\text{s}) e^{j \theta_{\text{RF},m}}
    \label{eq:TX_sync_PB}
\end{align}
The term $\theta_{\text{RF},m}$ represents a constant phase shift specific to chain $m$.
This shift arises from the cumulative phase response of amplifiers, switches, splitters, filters, and other front-end components.
As illustrated in \figref{fig:TX_sync_scheme}, the synchronization preamble has duration $t_\text{syn}=MNT_\text{s}$ and is repeated every $t_\text{obs}$ seconds to obtain phase observations for each chain periodically.
The locally received signal at the reference \ac{RF} chain is the sum of the transmit signals
\begin{align}
    r_\text{TX}(t) = \sum_{m=1}^{M} s_{\text{TX},m}(t).
\end{align}
Down-converting using the reference chain's local oscillator at frequency $f_\text{c}$ yields the baseband received signal
\begin{align}
    y_\text{BB}(t) = r_\text{TX}(t) e^{-j2\pi f_\text{c}t}.
\end{align}
Note that the received signal gain is neglected, assuming it is compensated by an automatic gain control (AGC).
Furthermore, the phase shift of the \ac{RF} components at the reference chain is neglected, as it will induce an identical phase shift for all $s_{\text{TX},m}(t)$ and only the relative phase relation of \ac{RF} chains are of interest.
For simplicity, the phase noise introduced by the oscillator of the reference chain is neglected under the assumption that $t_\text{syn}$ is sufficiently short.
As a frequency offset are already modeled at the transmitting \ac{RF} chain, the synthesized carrier frequency at the reference chain is assumed to be ideal.\\

Assuming $|x_\text{BB}(t)|^2=1$, the system function estimate for chain $m$ is obtained by multiplying the received baseband signal with the complex conjugate of the transmitted signal and windowing to the appropriate \ac{TDMA} slot:
\begin{align}
    \hat{h}_{\text{TX},m}(t) = y_\text{BB}(t) \cdot x_\text{BB}^*(t) \cdot \text{rect} \left[ \frac{t - \left( m + \frac{N}{2}\right)T_\text{s}}{NT_\text{s}} \right].
    \label{eq:tx_split_sync_blocks}
\end{align}
The time-dependent phase is then found as $\arg[\hat{h}_{\text{TX},m}(t)]$.
Assuming this phase consists of a constant term with additive Gaussian noise, as elaborated on in section~\ref{sec:phase_noise}, the optimal estimator is the time average \cite{Kay_Steven10_5555_151045}.
Averaging this phase over the synchronization interval yields the phase estimate
\begin{align}
    \hat{\theta}_m = \frac{1}{NT_\text{s}} \int_{0}^{NT_\text{s}} \arg[\hat{h}_{\text{TX},m}(t)]dt.
    \label{eq:phase_est_TX}
\end{align}
The transmission controller \figref{fig:TX_sync_setup} then uses these phase estimates to align all $M$ transmitting \ac{RF} chains.
To achieve coherent transmission, the baseband data signal transmitted from all chains $z[n]$ is pre-compensated with the most recent phase estimate
\begin{align}
        \tilde{z}_{m,l}[n] = z[n] \cdot e^{-j \hat{\theta}_{m,l}} \cdot \text{rect} \left[ \frac{n - \tau_{\text{TX},l}}{t_\text{dm}} \right], \label{eq:coherent_tx_precoding}
\end{align}
where $\hat{\theta}_{m,l}$ is the phase estimate for chain $m$ from the $l$-th calibration interval $[l-1,~l]t_\text{obs}$, and $\tau_{\text{TX},l} = t_\text{syn} + t_\text{dm}/2 + l t_\text{obs}$ centers the data block within the transmission window \figref{fig:TX_sync_scheme}.
Note that the rectangular expression is used to apply the phase estimate to the data signal block (blue element in \figref{fig:TX_sync_scheme}).
Then the corresponding quasi-coherent transmitted pass-band signal is
\begin{align}
    \tilde{s}_{l}(t) &= \sum_{m=1}^{M} \tilde{z}_{m,l}(t) e^{j(2\pi f_{\text{c},m}t + \theta_{\text{OS},m}(t)+ \theta_{\text{RF},m})} \\ 
    &= \sum_{m=1}^{M} z(t) e^{j(2\pi f_\text{c}t + \theta_m(t) -  \hat{\theta}_{m,l})} \cdot \text{rect}\left[\frac{t - \tau_{\text{TX},l}}{t_\text{dm}}\right],\nonumber
    \label{eq:coherent_tx_bp_signal}
\end{align}
with ${\theta_m(t) = 2\pi \Delta f_mt +\theta_{\text{OS},m}(t) + \theta_{\text{RF},m}}$ and ${\Delta f_m=f_{\text{c},m} -f_{\text{c}}}$.

\subsection{Phase Noise Model}
\label{sec:phase_noise}

IEEE defines phase noise as the random fluctuation in the phase of a periodic signal, typically characterized in the frequency domain by its spectral density \cite{9364950}.
In the time domain, this fluctuation manifests as timing error, or jitter.
The phase error resulting from the timing error at chain $m$ for an oscillator operating at frequency $f_\text{c}$ is $\theta_{\text{OS},m}(t) = 2\pi f_\text{c} \alpha_m(t)$.

In the case of a free running \ac{VCO}, the jitter can be represented by a Wiener process \cite{demir847872}.
In continuous time,
\begin{align}
\alpha_\text{VCO}(t) = \sqrt{c_\text{VCO}} \int_{0}^{t} \xi(t') dt',
\end{align}
where $c_\text{VCO}$ is the oscillator constant and $\xi(t)$ represents a Gaussian process with unit variance.
When this process is sampled at $T_\text{s}$ to obtain $\xi(i T_\text{s})$ discrete increments, the jitter can be written as
\begin{align}
    \alpha_\text{VCO}[n] = 
    \begin{cases}
        0,&n=0\\
        \sqrt{c_\text{VCO}T_\text{s}} \sum_{i=0}^{n-1} \xi(i T_\text{s}),& n>0
    \end{cases}.
\end{align}
Note that to obtain the Wiener process, the Gaussian process is scaled with the oscillator constant $c_\text{VCO}$.
This yields a $1/f^2$ phase noise spectrum with \SI{3}{dB} bandwidth $f_\text{3dB} = \pi f_\text{c}^2 c_\text{VCO}$.

In a \ac{PLL}, the \ac{VCO} is placed in a control loop that suppresses long-term drift by comparing its output to a stable reference signal.
It is assumed that both reference oscillator and \ac{VCO} can be modeled as free running oscillators and that the \ac{PLL} is in locked state.
Then the \ac{PLL} output jitter in discrete time is given by \cite{collmann2025practicalanalysisunderstandingphase}
\begin{align}
    \alpha_\text{PLL}[n] = 
    \begin{cases}
      0, & n = 0 \\
       \sum_{i=0}^{n-1} \left( \alpha_\text{PLL}[i] - \alpha_\text{REF}[i] \right)& \\
      \cdot \left( -2\pi f_\text{PLL}T_\text{s} \right) + \alpha_\text{VCO}[n-1], & n > 0
    \end{cases}.
    \label{eq:pn_pll}
\end{align}
The bandwidth of the \ac{PLL} is represented by $f_\text{PLL}$ and $c_\text{REF}$ refers to the oscillator constant of the reference oscillator.

The \ac{SDR} platforms used in our measurements employ \ac{PLL}'s locked to a common $\SI{10}{MHz}$ reference.
The model for a \ac{PLL} can be simplified under the following conditions:
\begin{enumerate}
    \item the oscillator constant for the reference oscillator is sufficiently high so that during an observation period $t_\text{obs}$ the condition $c_\text{VCO} >> c_\text{REF}$ applies, 
    \item the \ac{PLL} bandwidth is sufficiently high $f_\text{PLL} >> f_\text{3dB}$
    \item the process is sampled so that $\frac{1}{T_\text{s}} > f_\text{PLL}$,
    \item the observation time is short $t_\text{obs} < \frac{c_\text{VCO} - 3c_\text{REF}}{c_\text{REF} 2 \pi f_\text{PLL}}$ \cite{collmann2025practicalanalysisunderstandingphase}.
\end{enumerate}
Then, the jitter at \ac{PLL} output can be treated as white noise and described by $\alpha_\text{PLL}[n] = \alpha_0 + w[n]$, where $w[n] \sim \mathcal{N}(0, c_\text{VCO}T_\text{s})$ is white Gaussian noise.
The term $\alpha_{0} \sim \mathcal{U}(0, f_\text{c}^{-1})$ refers to a uniformly distributed constant time shift.
Compensating this constant offset is essential for initial phase alignment, while periodic calibration tracks the time-varying component $w[n]$.
In continuous time the phase of \ac{RF} chain $m$ is $\theta_m(t) = 2\pi f_\text{c} \alpha_{\text{PLL},m}(t)$.
This approximation is validated through measurements in the following section.

\section{Measurement Results}\label{sec:impairment_model}

\subsection{Measurement Setup}

\begin{table}[tb]
    \centering
    \begin{center}
        \caption{System parameters measurement.}
            \label{table:PN_parameters}
        \resizebox{.7\linewidth}{!}{
    \begin{tabular}{| l | l | l |} 
        \hline
        Parameter & Symbol & Value \\
        \hline
        Carrier frequency & $f_\text{c}$ & $\SI{3.75}{GHz}$\\
        Bandwidth & $B$ & $\SI{1}{MHz}$\\
        Sample frequency & $f_\text{s}$ & $\SI{4}{MHz}$\\
        TX RF chains & $M$ & $6$\\
        Observations & $L$ & $10000$\\
        Obs. Interv. & $t_\text{obs}$ & $\SI{100}{ms}$\\
        Samples & $N$ & $2500$\\
        \hline
    \end{tabular}
    }
    \end{center}
\end{table}

 \begin{figure}[tb]
     \centering   
    \includegraphics[width=\linewidth]{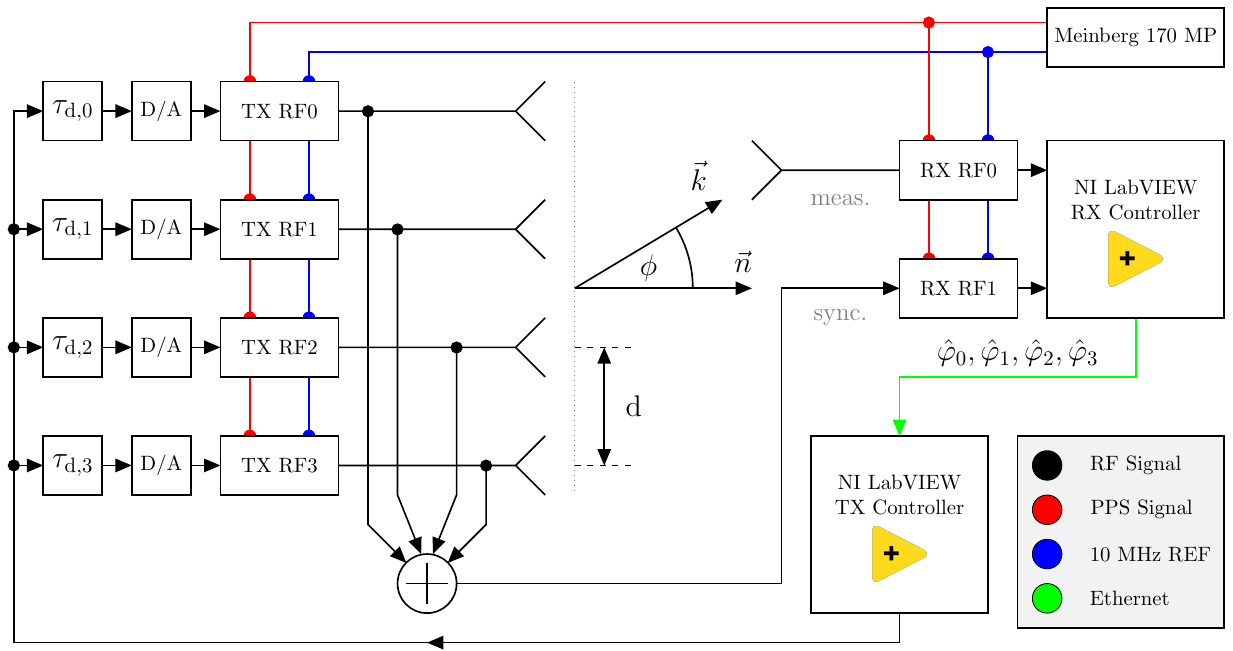}
    \caption{System setup for transceiver with $4$ TX and real-time calibration\cite{Coll202501_2}.}
    \label{fig:schematic_setup}
\end{figure} 
\begin{figure}[tb]
    \centering   
    \includegraphics[width=0.9\linewidth]{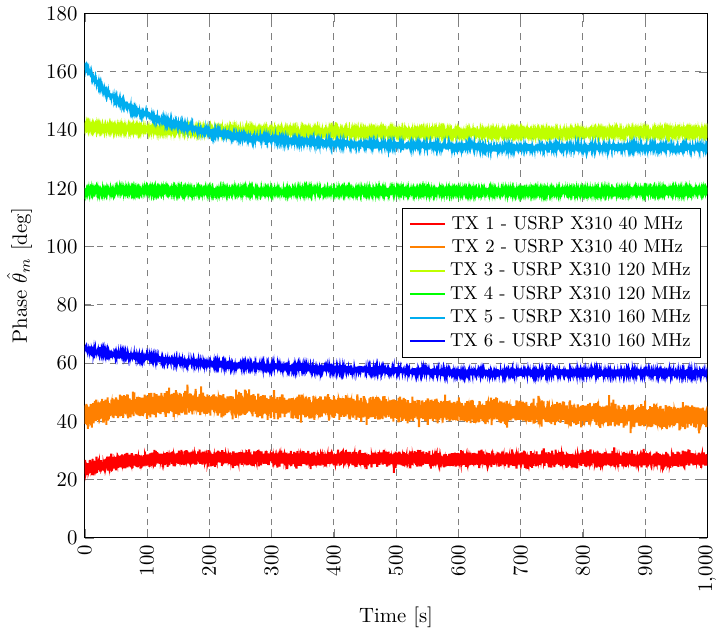}
    \caption{Estimated phases for $6$ transmit chains \cite{vnn4-cz93-26}.}
    \label{fig:TX_phase_meas}
\end{figure}

\figref{fig:schematic_setup} shows the measurement setup used to evaluate the proposed real-time phase calibration method.
The system follows the architecture described in \cite{Coll202501_2}.
Each TX \ac{RF} chain transmits the synchronization signal in a \ac{TDMA} scheme.
The synchronization signal is transmitted over cable to receiver chain \texttt{RF1} to isolate the phase noise of the transmit chains from channel-induced variations.
From this recorded signal, the RX controller estimates the phases of the transmitting \ac{RF} chains.
These phase estimates are sent to the TX controller on a host PC for precoding to allow for coherent transmission as shown in eq.~\eqref{eq:coherent_tx_precoding}.

Measurements were performed on six transmit chains using three different versions of the \ac{USRP} \texttt{X310}.
Key parameters for the measurement are given in Table~\ref{table:PN_parameters}.
With observation interval $t_\text{obs} = \SI{100}{ms}$ and $L = 10000$ observations, the total measurement duration was $\SI{1000}{s}$.
\figref{fig:TX_phase_meas} illustrates the estimated phases over time. 

Several observations can be made from the measurement data \cite{vnn4-cz93-26} shown in \figref{fig:TX_phase_meas}.
During the first $\SI{250}{s}$ of operation, the phases of TX5 and TX6 (cyan and blue traces) drift by approximately $\SI{25}{\degree}$ and $\SI{5}{\degree}$, respectively, despite all chains sharing a common $\SI{10}{MHz}$ reference.
This effect is likely caused by temperature-induced changes in the phase response of the \ac{RF} frontend during warm up.
Other chains such as TX3 and TX4 (lime and green trace) and exhibit negligible drift during the measurement.
This is likely due to these chains being already warm from previous measurements when the recording began.
These observations highlight the importance of allowing sufficient warm up time of \ac{RF} components in practical deployments.
As demonstrated in the following subsections, the proposed calibration method is suitable for tracking these phase variations, enabling coherent transmission despite this temperature induced drift.


\begin{figure*}[tb]
    	\centering                      
         \begin{minipage}{0.32\textwidth}
             \centering   
        	\includegraphics[width=\textwidth]{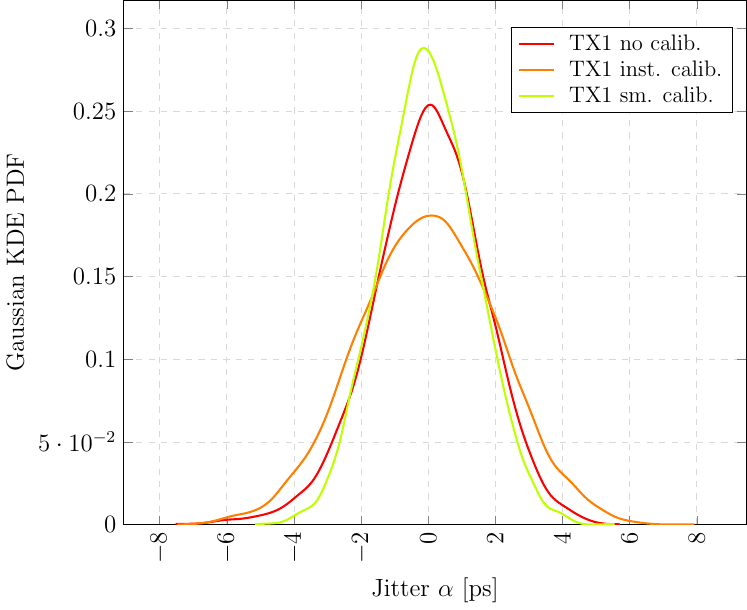}
            \caption{\ac{PDF} for TX1 before/after calibration}
        	\label{fig:PDF_KDE_6TX_no_avg_TX1}
        \end{minipage} 
         \begin{minipage}{0.30\textwidth}
             \centering   
            \includegraphics[width=\textwidth]{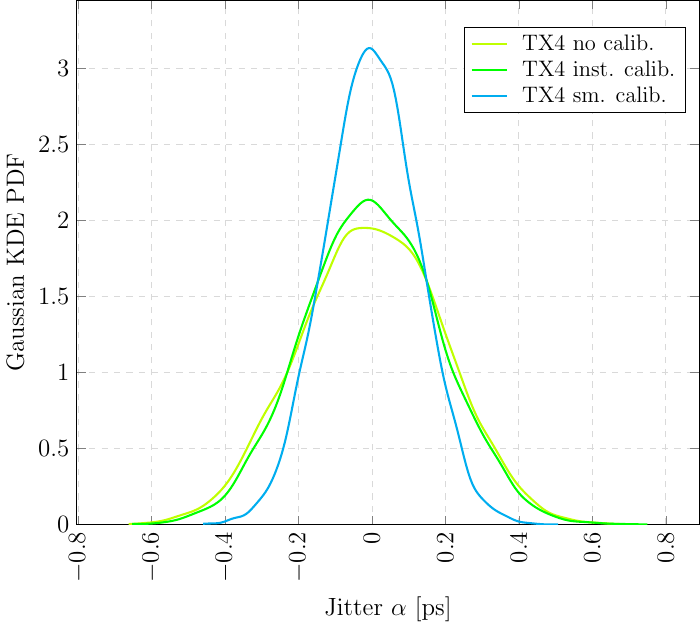}
            \caption{\ac{PDF} for TX4 before/after calibration}
        	\label{fig:PDF_KDE_6TX_no_avg_TX4}
        \end{minipage} 
        \begin{minipage}{0.3\textwidth}
            \centering   
            \includegraphics[width=\textwidth]{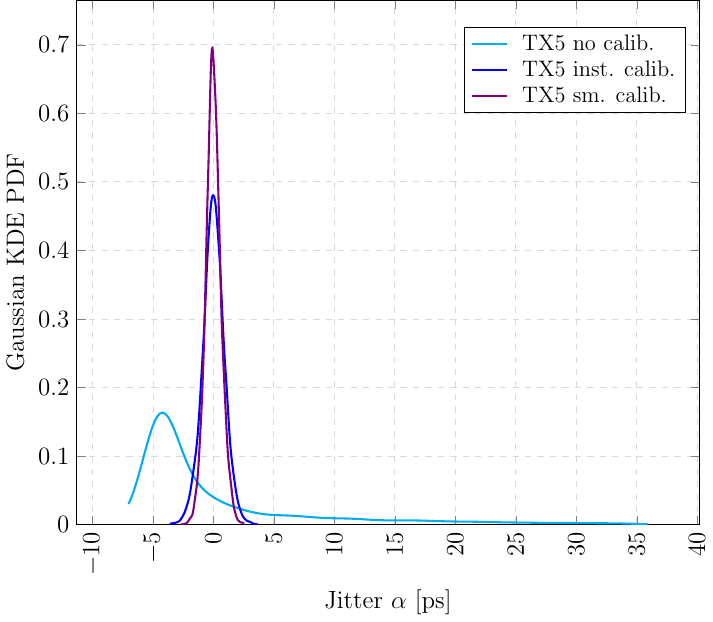}
            \caption{\ac{PDF} for TX5 before/after calibration}
        	\label{fig:PDF_KDE_6TX_no_avg_TX5}
        \end{minipage}
\end{figure*}

\subsection{Jitter distribution}


Analyzing the \ac{PDF} of the timing jitter provides insight into the effectiveness of the proposed calibration methods.
The jitter for chain $m$ at observation interval $l$ is obtained from the phase estimates as
\begin{align}
\alpha_{m,l} = \frac{\hat{\theta}_{m,l}}{2\pi f_\text{c}}.
\end{align}
The \ac{PDF} is estimated using a Gaussian \ac{KDE} applied to the measured jitter values.

Two calibration approaches are considered:
\begin{itemize}
    \item \textbf{Instantaneous calibration:} precoding with most recent phase estimate as in Eq.~\eqref{eq:coherent_tx_precoding}
    \item \textbf{Smoothed calibration:} precoding using the average of the last 10 phase estimates, $\frac{1}{10}\sum_{k=0}^{9} \hat{\theta}_{m,l-k}$, to reduce estimation noise.
\end{itemize}

As a quantitative metric, the \ac{RMS} cycle-to-cycle jitter is used with definition
\begin{align}
    \tau_{\text{RMS},m} = \frac{1}{2\pi f_\text{c}} \sqrt{\frac{1}{L-1} \sum_{l=0}^{L-1}|\hat{\theta}_{m,l+1} - \hat{\theta}_{m,l}|^2 }.
    \label{eq:rms_jitter}
\end{align}
Note that this metric is valid only when the phase is sufficiently stationary over the observation period \cite{loehning8257361325}.

Fig.~\ref{fig:PDF_KDE_6TX_no_avg_TX1}, \ref{fig:PDF_KDE_6TX_no_avg_TX4}, and \ref{fig:PDF_KDE_6TX_no_avg_TX5} show the jitter \ac{PDF}'s for TX1, TX4 and TX5, illustrating different time-dependent phase responses. Table~\ref{table:jitter_summary} summarizes the corresponding RMS jitter values.

\begin{table}[tb]
    \centering
    \begin{center}
        \caption{\ac{RMS} cycle-to-cycle jitter for different TX.}
            \label{table:jitter_summary}
        \resizebox{.7\linewidth}{!}{
    \begin{tabular}{| l | l | l | l |} 
        \hline
        &No calib. &Inst. calib. &Sm. calib.\\
        \hline
        TX1 & $\SI{1.66}{ps}$& $\SI{2.06}{ps}$& $\SI{1.39}{ps}$\\
        TX2 & $\SI{3.03}{ps}$& $\SI{3.20}{ps}$& $\SI{2.15}{ps}$\\
        TX3 & $\SI{728}{fs}$& $\SI{199}{fs}$ & $\SI{134}{fs}$\\
        TX4 & $\SI{193}{fs}$& $\SI{183}{fs}$ & $\SI{124}{fs}$\\
        TX5 & $\SI{7.78}{ps}$& $\SI{891}{fs}$ & $\SI{632}{fs}$\\
        TX6 & $\SI{2.99}{ps}$& $\SI{893}{fs}$ & $\SI{630}{fs}$\\
        \hline
    \end{tabular}
    }
    \end{center}
\end{table}

\subsubsection{RF Chains with Negligible Drift}

TX1 \figref{fig:PDF_KDE_6TX_no_avg_TX1} exhibits negligible drift during the total measurement duration, with a near-Gaussian jitter distribution (red) and \ac{RMS} value of $\SI{1.66}{ps}$.
Instantaneous calibration (orange) slightly increases the \ac{RMS} jitter to $\SI{2.06}{ps}$.
This effect occurs because the calibration correction applied in eq.~\eqref{eq:coherent_tx_bp_signal} effectively differentiates the phase, which amplifies high-frequency noise.
The increase is small relative to the uncalibrated jitter, indicating that calibration is not necessary when drift is negligible.
Smoothed calibration (lime) reduces the jitter to $\SI{1.39}{ps}$.

TX4 \figref{fig:PDF_KDE_6TX_no_avg_TX4} shows low intrinsic jitter ($\SI{193}{fs}$ \ac{RMS}).
Instantaneous calibration yields marginal improvement ($\SI{183}{fs}$ \ac{RMS}), while smoothed calibration achieves $\SI{124}{fs}$ \ac{RMS}.

\subsubsection{RF Chains with Significant Drift}

TX5 \figref{fig:PDF_KDE_6TX_no_avg_TX5} exhibits significant drift, as evident from its \ac{PDF} without calibration (cyan, $\SI{7.78}{ps}$ \ac{RMS}).
Instantaneous calibration effectively compensates this drift, resulting in a Gaussian distribution (blue, $\SI{891}{fs}$ \ac{RMS}).
Smoothed calibration further reduces jitter to $\SI{632}{fs}$ \ac{RMS} (violet), demonstrating the benefit of averaging.

After calibration, the \ac{RMS} jitter is consistent within each \ac{USRP} variant: TX3 and TX4 ($124$ to $\SI{134}{fs}$) and TX5 and TX6 ($630$ to $\SI{632}{fs}$) show excellent agreement. TX1 and TX2 ($1.39$ to $\SI{2.15}{ps}$) exhibit more variation but remain within the same order of magnitude.

\begin{figure}[tb]
\centering
\includegraphics[width=0.9\linewidth]{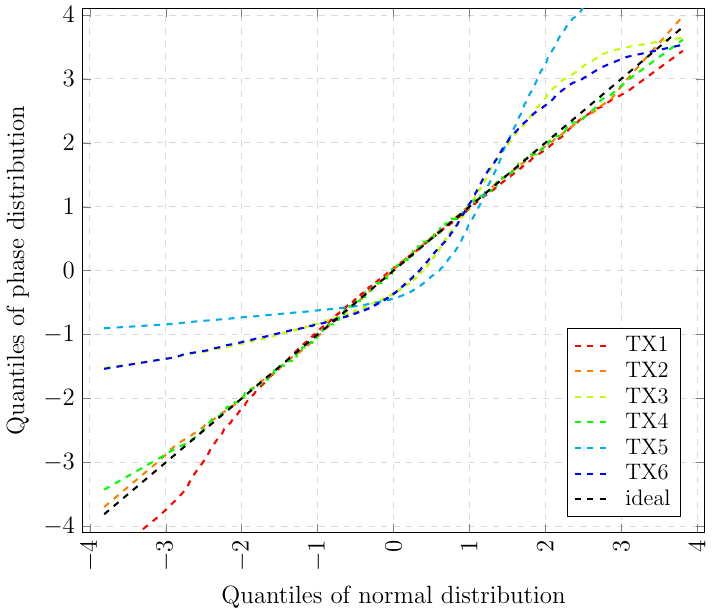}
\caption{Measured quantiles of phase plotted over quantiles of normal distribution}
\label{fig:QQ_plot_6TX}
\end{figure}
\begin{figure}[tb]
\centering
\includegraphics[width=0.9\linewidth]{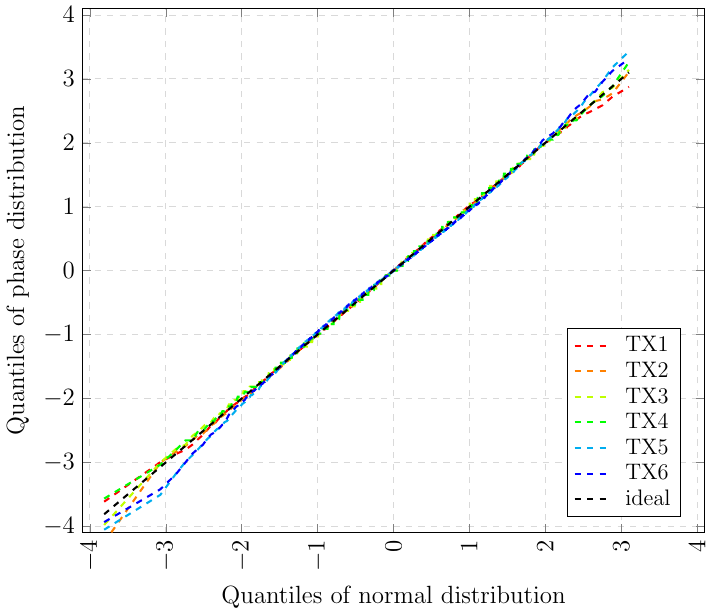}
\caption{Quantiles of phase after calibration plotted over quantiles of normal distribution}
\label{fig:QQ_plot_6TX_cmp}
\end{figure}

\subsection{Quantile-Quantile Plot}

A quantile-quantile (Q-Q) plot is a graphical method for comparing the similarity of a distribution to the normal distribution.
In this plot, the quantiles of the evaluated distribution (typically on the y-axis) are plotted against the quantiles of a normal distribution (typically on the x-axis).
The distributions are normalized by their standard deviations to allow for even comparison.
If a distribution function matches the normal distribution, the Q-Q plot data points fall on a straight line (see \figref{fig:QQ_plot_6TX} black trace).



\figref{fig:QQ_plot_6TX} shows the Q-Q plot for the estimated phases before calibration.
The distributions for TX3, TX5, and TX6 deviate substantially from the reference line, indicating a non-Gaussian \ac{PDF}.
This is consistent with \figref{fig:TX_phase_meas}, which shows significant drift for TX5 and TX6 over the measurement duration.
From \figref{fig:PDF_KDE_6TX_no_avg_TX5} it can also be seen that the \ac{PDF} for TX5 (cyan) is not Gaussian and that the phase of this chain drifts during the measurement.
For the other \ac{RF} chains TX1, TX2 and TX4 the distributions are approximately Gaussian.
In contrast, TX1, TX2, and TX4 are approximately Gaussian, though TX1 exhibits slight deviation at the tails of its distribution at higher quantiles.

After applying the proposed calibration, the Q-Q plot in \figref{fig:QQ_plot_6TX_cmp} shows that all distributions align closely with the normal distribution.
This indicates that the residual jitter and corresponding phase errors are Gaussian distributed, which suggests that the calibration has effectively whitened the phase noise.
A slight deviation at the tails of the distribution ($\pm 2$ standard deviations) is still visible.
However, this affects only a small fraction of the distribution.
Within the central $95\%$ of the probability mass, the fit to a normal distribution is decent.

\section{Simulation Results}\label{sec:evaluation}

Numerical simulations are performed to complement the hardware measurements by allowing systematic variation of the observation interval $t_\text{obs}$, enabling investigation of calibration performance beyond the fixed conditions of the experimental setup.

\subsection{Effect of Calibration in Time Domain}

\begin{figure*}[tb]
    	\centering                      
         \begin{minipage}{0.32\textwidth}
    	\centering
    	\includegraphics[width=\textwidth]{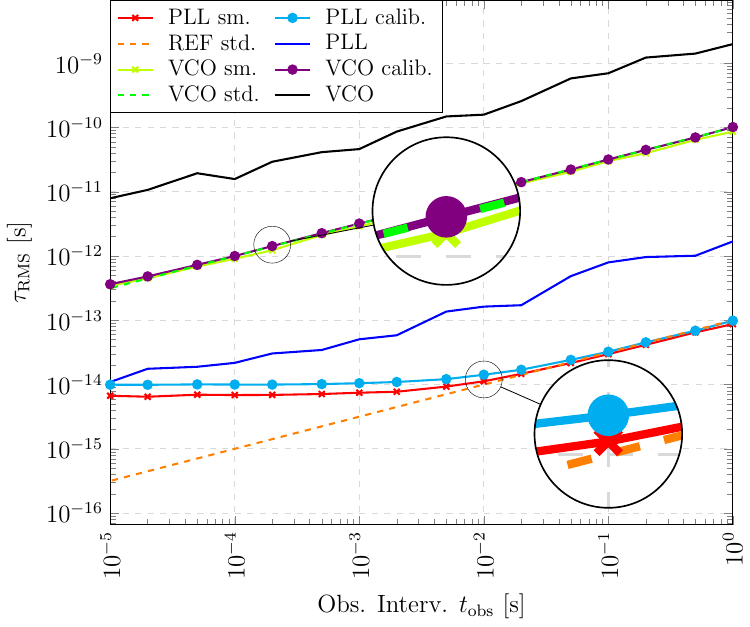}
        \caption{\ac{RMS} of Cycle-to-cycle jitter for different observation intervals.}
    	\label{fig:tau_rms}
        \end{minipage}
        \begin{minipage}{0.31\textwidth}
    	\centering
    	\includegraphics[width=\textwidth]{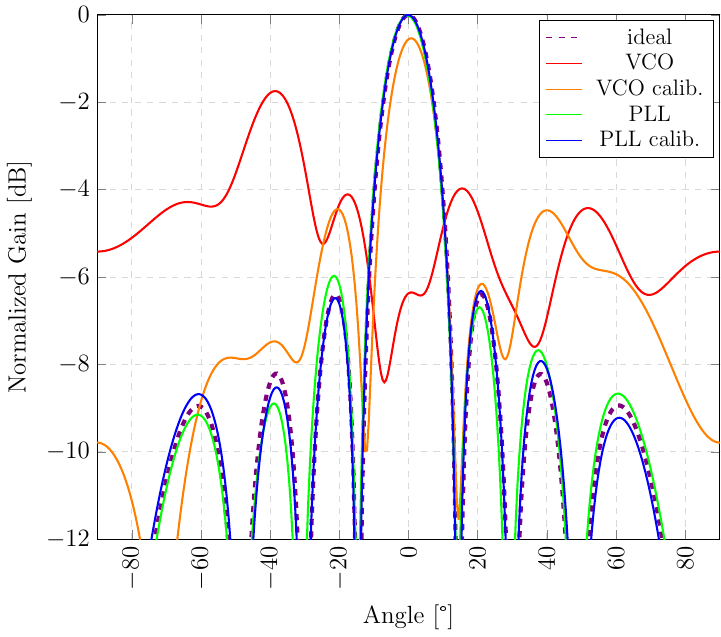}
        \caption{Array response under phase impairments with and without periodic calibration}
    	\label{fig:arr_resp}
        \end{minipage}
        \begin{minipage}{0.31\textwidth}
            	\centering
            	\includegraphics[width=\textwidth]{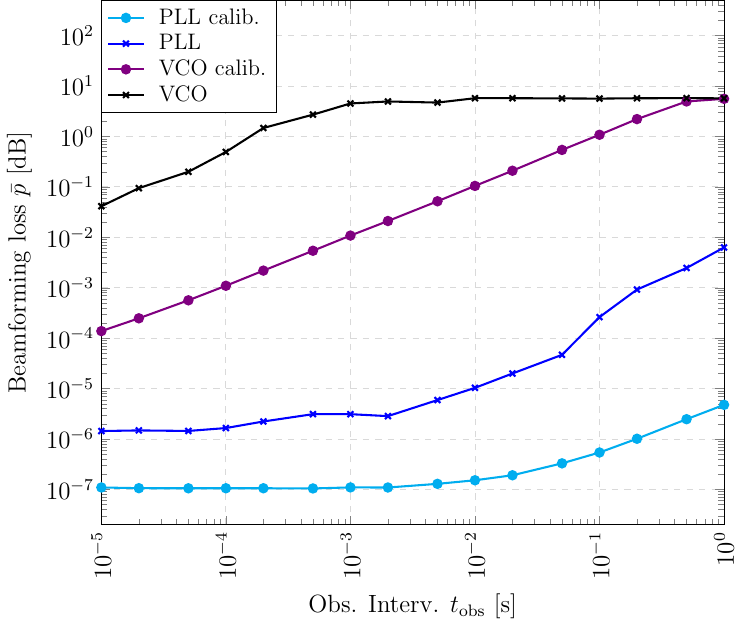}
                \caption{Average beamforming loss in steering direction for different observation intervals}
            	\label{fig:bf_loss}
        \end{minipage}
\end{figure*}


When continuous calibration is applied, the observation interval $t_\text{obs}$ has a direct effect on the residual jitter.
This section investigates this relationship for both free-running \ac{VCO} and a \ac{PLL}, with \ac{RMS} of the cycle-to-cycle jitter $\tau_\text{RMS}$ as defined in eq.~\eqref{eq:rms_jitter} as a metric.
Key simulation parameters are listed in Table~\ref{table:system_parameters_sim}.

\figref{fig:tau_rms} shows the \ac{RMS} of the jitter processes for the uncalibrated and calibrate case, at different observation intervals.
Two theoretical lower bounds are also shown: $\sqrt{c_\text{VCO} t_\text{obs}}$ for the \ac{VCO} and $\sqrt{c_\text{REF} t_\text{obs}}$ for the \ac{PLL}, representing the fundamental limits set by oscillator properties.

\subsubsection{Free-Running VCO}

For the uncalibrated \ac{VCO} (black trace), $\tau_\text{RMS}$ increases with $t_\text{obs}$, consistent with the variance scaling $\sigma^2 = c_\text{VCO} t$ \cite{collmann2025practicalanalysisunderstandingphase}.
Both instantaneous and smoothed calibration (lime and violet traces) achieve the theoretical lower bound $\sqrt{c_\text{VCO} t_\text{obs}}$ across all observation intervals, confirming that the calibration method optimally compensates the \ac{VCO} phase noise.

\subsubsection{Phase-Locked-Loop}

The uncalibrated \ac{PLL} (blue trace) exhibits a similar behaviour, with a increase of $\tau_\text{RMS}$ over $t_\text{obs}$.
After calibration, three observations stand out. 
First, smoothed calibration consistently outperforms the instantaneous calibration by a slight margin.
Second, the calibrated jitter \ac{RMS} for longer intervals is asymptotically bounded by the reference oscillator limit $\sqrt{c_\text{REF} t_\text{obs}}$ (orange trace).
Third, below approximately $\SI{10}{ms}$, the jitter reaches a floor and does not improve with shorter $t_\text{obs}$.
This however is not due to a limitation on the calibration but innate properties of a \ac{PLL}.
In the transition region between \ac{VCO}-dominated and reference oscillator-dominated regimes, the \ac{PLL} jitter variance is bounded by \cite{collmann2025practicalanalysisunderstandingphase}
\begin{align}
\sigma^2 = \frac{c_\text{VCO} - 3c_\text{REF}}{4\pi f_\text{PLL}}.
\end{align}

For the given parameters in Table~\ref{table:system_parameters_sim} this evaluates to a variance of $\SI{7.957e-28}{s^2}$, with \ac{RMS} jitter of $\SI{2.821e-14}{s}$.
The smoothed calibration trace in \figref{fig:tau_rms} reaches exactly this value, demonstrating that the proposed method achieves the fundamental performance bound for \ac{PLL}'s.

\subsection{Effect of Calibration on Beamforming}




On the transmit side, coherence of the \ac{RF} chains in a fully digital array is essential to allow for beamforming.
Consequently, significant imperfections in phase calibration of the array will result in a loss in beamforming gain in the desired steering direction.

\figref{fig:arr_resp} shows the array response at boresight for arrays affected by \ac{VCO} and \ac{PLL} phase impairments, assuming an initial calibration has been performed, with and without subsequent periodic calibration.
Without periodic calibration, the array subjected to phase noise from a \ac{VCO} (red trace) shows no coherent beamforming pattern.
With periodic calibration (orange trace), a clear main lobe emerges.
A small offset from the ideal gain remains, as the \ac{VCO} jitter over the $\SI{100}{ms}$ observation interval is still significant.

Since the \ac{PLL} output process does not drift significantly for the considered observation interval, the array response even without applying periodic calibration (green trace) is already near ideal.
Applying the calibration procedure to compensate the \ac{PLL} jitter process (blue trace) yields only a small improvement in the side lobes.
For the \ac{PLL} the cases with and without periodic calibration achieve near-optimal beamforming gain.
This is because the \ac{PLL} holds its output phase nearly constant over the $\SI{100}{ms}$ observation interval, exhibiting negligible drift.


As a metric to assess the beamforming performance with and without periodic calibration, the average beamforming loss in steering direction is considered.
This average beamforming loss $\bar{p}$ is calculated as the difference between ideal gain and obtained gain in steering direction and averaged over steering angles $\phi \sim \mathcal{U} (\SI{-60}{\degree},\SI{60}{\degree})$.

\figref{fig:bf_loss} shows the average beamforming loss for \ac{VCO} and \ac{PLL} cases with and without periodic calibration.
For the considered array size, a maximum gain of $\SI{6}{dB}$ can be achieved.
This value corresponds to the loss observed for the \ac{VCO} without periodic calibration (black trace) at longer observation intervals.
Below an observation interval of $\approx \SI{1}{ms}$, the loss declines.
This is due to the observation interval being sufficiently short that the \ac{VCO} cannot drift that far in this time period, allowing the initial calibration to remain partially effective.
When the \ac{VCO} jitter is periodically calibrated for (violet trace) but the observation interval is long ($\SI{1}{s}$), there is no improvement over the case with only initial calibration.
This is due to the \ac{VCO} drifting sufficiently far between calibration times, rendering beamforming ineffective.
For all shorter observation intervals, periodic calibration yields a significant improvement in beamforming loss.

In the case of a \ac{PLL} with only initial calibration (blue trace), the beamforming loss is negligibly small meaning that there is no need for periodic calibration with the given choice of parameters.
Still, periodic calibration (cyan trace) can further reduce this loss by approximately an order of magnitude.
For short observation intervals below $\SI{1}{ms}$ the beamforming loss reaches a floor.
This corresponds to the fundamental \ac{PLL} jitter limit as discussed previously and observed also in \figref{fig:tau_rms}.




\begin{table}[tb]
    \centering
    \begin{center}
        \caption{System parameters simulation.}
        \vspace{-2mm}
            \label{table:system_parameters_sim}
        \resizebox{0.7\linewidth}{!}{
        \begin{tabular}{| l | l | l |} 
         \hline
         Parameter & Symbol & Value \\
         \hline
         Carrier frequency & $f_c$ & $\SI{3.75}{GHz}$ \\ 
         Sample frequency & $f_\text{s}$ & $\SI{20}{MHz}$ \\
         TX RF chains & $M$ & $8$\\
         Observations & $L$ & $1000$\\
         Obs. Interv. & $t_\text{obs}$ & $\{\SI{1}{ms}, \dots, \SI{10}{\micro s} \}$\\
         Samples & $N$ & $100$\\
         Osc. Const. VCO & $c_\text{VCO}$ &$\SI{1e-20}{s}$ \\
         Osc. Const. REF & $c_\text{REF}$ &$\SI{1e-26}{s}$\\
         PLL Bandw. & $f_\text{PLL}$ &$\SI{1}{MHz}$\\         
         \hline
        \end{tabular}
    }
    \end{center}
    \vspace{-5mm}
\end{table}

\section{Conclusion}\label{sec:conclusions}

Commonly known methods for compensation of phase noise in \ac{MIMO} systems typically apply phase noise calibration at the receiver.
However, transmit-side beamforming with large arrays requires real-time phase coherence across all \ac{RF} chains during transmission. 
This paper has presented and validated a simple method for real-time phase calibration of fully digital transmit arrays.

Two calibration approaches were compared: instantaneous compensation using the most recent phase estimate, and smoothed compensation using a time average of the last 10 estimates.
Measurements on six \ac{USRP} X310 chains demonstrated that both methods significantly reduce phase drift and jitter.
Key results include:
\begin{itemize}
\item \ac{RMS} cycle-to-cycle jitter reduced to as low as $\SI{124}{fs}$ for \ac{PLL}'s, with smoothed calibration consistently outperforming instantaneous calibration.
\item Thermal drift of up to $\SI{25}{\degree}$ observed during warm-up period, despite a common $\SI{10}{MHz}$ reference, highlighting the need for calibration even with shared reference oscillator.
\item After calibration, residual phase errors are Gaussian distributed, as confirmed by QQ plot analysis, indicating whitening of the phase noise.
\item Simulation results showed that the proposed method achieves the fundamental \ac{PLL} jitter limit of $\SI{2.28e-14}{s}$ for the given system parameters and optimal choice of observation interval duration.
\end{itemize}


These results demonstrate that the proposed low-complexity calibration methods are well-suited for synchronizing \ac{SDR}-based testbeds and ensuring coherence in \ac{MIMO} channel measurements.

Future work could extend this approach to joint transmitter-receiver calibration, integrate it with over-the-air reciprocity calibration methods, and investigate its performance under mobile conditions where Doppler effects become significant.
	\section*{Acknowledgment}

This work was supported by BMBF under the project KOMSENS-6G (16KISK124).
    \bibliographystyle{IEEEtran}
	\bibliography{references}

\end{document}